\documentclass[12pt,preprint]{aastex}

\usepackage{emulateapj5}

\lefthead{Dewangan, Griffiths, \& Schurch}
\righthead{\xmm{} Observations of MCG-5-23-16}
\slugcomment{Revision 1, Submitted to ApJ}

\newcommand\asca{{\it ASCA}}
\newcommand\ginga{{\it Ginga}}
\newcommand\sax{{\it BeppoSAX}}
\newcommand\chandra{{\it Chandra}}

\newcommand\rxte{{\it RXTE}}
\newcommand\xmm{{\it XMM-Newton}}

\newcommand\ks{{\rm~ks}}
\newcommand\kev{{\rm~keV}}
\newcommand\ev{{\rm~eV}}
\newcommand\kms{\ifmmode {\rm~km\ s}^{-1} \else ~km s$^{-1}$\fi}
\newcommand\Hunit{\ifmmode {\rm~km\ s}^{-1}\ {\rm Mpc}^{-1}
        \else ~km s$^{-1}$ Mpc$^{-1}$\fi}
\newcommand\ctssec{\ifmmode {\rm~count\ s}^{-1} \else ~count s$^{-1}$\fi}
\newcommand\ergsec{\ifmmode {\rm~erg\ s}^{-1} \else
        ~erg s$^{-1}$\fi}
\newcommand\funit{\ifmmode {\rm~erg\ s}^{-1}\;{\rm cm}^{-2} \else
        ~ergs s$^{-1}$ cm$^{-2}$\fi}
\newcommand\phflux{\ifmmode {\rm~photon\ s}^{-1}\;{\rm cm}^{-2}
        \else   ~photon s$^{-1}$ cm$^{-2}$\fi}
\newcommand\efluxA{\ifmmode {\rm~erg\ s}^{-1}\;{\rm cm}^{-2}\;{\rm
        \AA}^{-1} \else ~erg s$^{-1}$ cm$^{-2}$ \AA$^{-1}$\fi}
\newcommand\efluxHz{\ifmmode {\rm~erg\ s}^{-1}\;{\rm cm}^{-2}\;{\rm
        Hz}^{-1} \else ~erg s$^{-1}$ cm$^{-2}$ Hz$^{-1}$\fi}
\newcommand\cc{\ifmmode {\rm~cm}^{-3} \else cm$^{-3}$\fi}
\newcommand\FWHM{\ifmmode {\rm~FWHM} \else ${\rm~FWHM}$\fi}
\newcommand\Msun{\ifmmode M_{\odot} \else $M_{\odot}$\fi}
\newcommand\Lsun{\ifmmode L_{\odot} \else $L_{\odot}$\fi}

\newcommand\hbeta{\ifmmode {\rm H}\beta \else H$\beta$\fi}
\newcommand\Kalpha{\ifmmode {\rm K}\alpha \else K$\alpha$\fi}
\newcommand\nh{\ifmmode N_{\rm H} \else N$_{\rm H}$\fi}
\usepackage{graphicx}
\usepackage{here}

\received{2002 August 28}
\begin{document}

\title{\xmm{} observations of Broad FeK$\alpha$ emission from a Seyfert~1.9 galaxy MCG-5-23-16} 

\author{G. C. Dewangan, R. E. Griffiths, and N. J. Schurch} 
\affil{Department of Physics, Carnegie Mellon University, 5000 Forbes Ave, Pittsburgh, PA 15213 USA}
\email{gulabd@cmu.edu}

\begin{abstract}
\xmm{} observations of the bright Seyfert~1.9 galaxy MCG-5-23-16 have
revealed a broad FeK$\alpha$ emission line  which is nearly
symmetric  
 in contrast to the broad and red-shifted asymmetric
FeK$\alpha$ line sometimes observed from Seyfert~1 galaxies.  The
FeK$\alpha$ line has 
two distinct components -- a narrow unresolved
component with equivalent width of $\sim40\ev$ and a broad component
with full width at half maximum of $\sim40000\kms$ and equivalent
width of $\sim120\ev$. An absorption feature  at $\sim7.1\kev$ has also been
observed.  The energies of the emission and absorption features are
consistent with those arising from neutral iron.
The broad component is consistent with an FeK$\alpha$ emission line 
expected from a relativistic accretion disk around a Schwarzschild or 
a Kerr black hole. Alternatively, most of the flux in the broad 
component could also be modeled as reflection emission which mimics 
emission line like feature due to the presence of iron K-shell edge at 
$\sim7.1\kev$, however, the reflection fraction, $R\sim3$, is much higher than that inferred from the \sax{} observations ($R\sim0.5$). 
The disk inclination angle of $\sim 47\deg$, inferred from the
 disk-line fits, and the absorption column ($\nh\sim
 10^{22}{\rm~cm^{-2}}$), inferred from the low-energy spectral
 curvature due to photoelectric absorption, suggest that our line of
 sight passes through the outer edge of a putative torus and are
 consistent with those expected for a Seyfert~1.9 galaxy falling
 within the unification scheme. 
The strength of the narrow iron K$\alpha$ emission and the
 optical depth of the iron K absorption edge suggest their origin in
 the putative torus with $\nh\sim 10^{24}{\rm~cm^{-2}}$ in the inner
 regions and $\nh \sim 10^{22}{\rm~cm^{-2}}$ in the outer edges. 
 The
 strength of the broad component of FeK$\alpha$ varied by a factor of
 $\sim 2$ between the two \xmm{} observations taken $\sim
 6{\rm~months}$ apart, while the narrow component of FeK$\alpha$ and
 the continuum flux did not appear to vary appreciably.  There is
 evidence for a weakening in the strength of the broad iron K$\alpha$
 emission with the flattening of the observed continuum. This can
 perhaps be explained if the shape of the continuum is coupled with
 the ionization stage of the reflector.
\end{abstract}

\section{Introduction}
The Seyfert-type active galactic nuclei (AGNs) appear to be
 intrinsically X-ray loud and the variation in their X-ray emission is
 largely due to different absorbing columns along the respective lines
 of sight. The type~1 and type~2 Seyferts and the intermediate
 Seyferts (types 1.5,1.8, and 1.9), all show a similar form in their
 intrinsic X-ray emission -- a power law with photon index of $\sim
 1.9$ which can extend up to a few$\times100\kev$, an FeK$\alpha$ line
 at $\sim6.4\kev$, and a reflection hump in the $\sim 10-100\kev$
 region. However, the obscured Seyferts show a low-energy cut-off of
 the intrinsic power law due to photoelectric absorption. The exact
 energy of the cut-off depends on the depth of the absorption
 column. The obscured Seyferts also show unabsorbed soft X-ray
 emission composed of a number of photo-excited emission lines \citep{Kinkhabwalaetal02,Schurchetal03}. This
 component is often extended, non-variable and cannot be the emission
 from the accretion disk. X-ray emission arising from the accretion
 disks of obscured Seyfert galaxies (e.g., the broad iron K$\alpha$
 emission) provide us with the opportunity to study the geometry and
 physical conditions in the inner regions surrounding a super massive
 black hole (SMBH), thus helping us in understanding the diversity in
 the AGN characteristics.

The iron K$\alpha$ emission is the most prominent ubiquitous 
feature in the $2-10\kev$ region of the X-ray spectra of AGNs.  
This emission appears to be a
 general characteristic of AGNs. Complex iron K$\alpha$ emission
 consisting of multiple components and/or asymmetric profiles arising
 from neutral and/or ionized iron have been observed from many
 Seyfert~1 galaxies and provide the strongest evidence for the
 existence of cold material in the central regions of AGNs. The broad
 iron K$\alpha$ emission is the key feature in studies of the geometry and
 nature of the accretion disks around black holes. The line profile
 can also be used to infer the Schwarzschild or Kerr nature of black
 holes.  While some Seyfert~1 galaxies show broad iron K-shell
 emission, there are only a few Seyfert~2 galaxies with broad iron
 K$\alpha$ emission. This is likely due to the difficulty in detecting
 the broad wings in the presence of large absorbing columns along our
 line of sight to a type 2 nucleus.  The narrow-emission line galaxies
 (NELGs) are good targets for study of the iron K$\alpha$ emission from the
 obscured AGNs. Optically, NELGs are the same as Seyfert~2s and show
 narrow optical emission lines, often with  broad wings of H$\alpha$ and
 P$\beta$ which makes them Seyfert~1.9 nuclei.  
 NELGs are bright and variable X-ray sources which were discovered in
 early X-ray surveys \citep{Marshalletal79, Griffithsetal78}.  In this paper, we
 present a detailed study of the iron K$\alpha$ emission from an
 obscured Seyfert galaxy MCG-5-23-16 using \xmm{} observations.

MCG-5-23-16 is an S0 galaxy \citep{FWM00} first discovered in X-rays
by \citet{Schnopperetal78} who also found its' optical spectrum to be
that of a narrow emission-line galaxy at a redshift of
$0.0083$. \citet{Veronetal80} classified this galaxy as a Seyfert~1.9
based on the possible presence of a broad component in
H$\alpha$.  MCG-5-23-16 with a V-magnitude of $13.7$ is one of the
brightest Seyfert galaxies in hard X-rays and is also one of the few
Seyferts detected at $\gamma$-ray energies
\citep{Bignamietal79,Pollocketal81}. Its $2-10\kev$ flux,
measured to be $\sim 8\times10^{-11}{\rm~erg~cm^{-2}~s^{-1}}$ in 1978
\citep{Tennant83}, decreased by a factor of four in 1989
\citep{NP94}, and then increased to
$\sim9\times10^{-11}{\rm~erg~cm^{-2}~s^{-1}}$ in 1996 \citep{WKP98}
before decreasing again to $7\times10^{-11}{\rm~erg~cm^{-2}~s^{-1}}$ in 2001
(this work).  Previous X-ray observations of MCG -5-23-16 showed a
highly absorbed X-ray spectrum
($\nh\sim10^{22}{\rm~cm^{-2}}$,\citealt{TP89}) and a strong iron
K$\alpha$ fluorescence line with equivalent width of $300-400\ev$
\citep{SRV92,NP94,Weaveretal97,WKP98}. Weaver et al. (1997) reported a
complex iron K-shell line profile in an \asca{} observation of this
source. These authors discovered a narrow core at the systemic
velocity of the galaxy and wings to the red and blue sides of this
core. They modeled the line profile with three Gaussian having
central energies $5.37$, $6.37$, and $6.58\kev$. Alternatively, the data could
also be modeled with a line profile from an accretion disk,
superimposed upon a narrow Gaussian profile.  These data have provided some
of the best evidence to date for contributions from two distinct
regions to the observed iron line profile.

In this paper we present the first results from the \xmm{}
observations of MCG-5-23-16. The paper is organized as follows. In
Sect.~\ref{obs}, we describe the \xmm{} observations and the data
selection. In Sect.~\ref{analysis}, we present the detailed spectral
analysis of the FeK$\alpha$ region. We discuss our results in
Sect.~\ref{discuss} followed by conclusions in Sect.~\ref{conclusion}.

\section{Observations and Data selection \label{obs}}

\xmm{} observatory \citep{Jansenetal01} has three Wolter type 1 X-ray telescopes with three European Photon Imaging Cameras (EPICs) -- one PN \citep{Struderetal01} and two MOS CCD \citep{Turneretal01} cameras as the imaging spectrometers. 
 All \xmm{} instruments 
operate simultaneously. \xmm{} observed MCG-5-23-16 twice on 13 May 2001 
and 1 December 2001 for 38ks and 25ks, respectively. The EPIC PN and MOS 
observations were carried out 
in the full frame mode  using the medium filter. 
Here we present the data from the PN and MOS cameras.

The raw PN and MOS events were processed and filtered using the most recent 
updated calibration database and analysis software ({\tt SAS v5.3.3}) available 
in December 2002. Events in the bad pixels file and those adjacent pixels were discarded.
Events with pattern 0-4 (single and double) for the PN and 0-12 (similar to \asca{} 
event grades 0-4) for the MOS cameras were selected for both the observations of 
2001. 
Examination of background light curves extracted from source-free regions 
showed that both the observations were affected by particle induced flares. 
These flares are characterized by strong and rapid variability in the background 
light curve. The particle induced events were filtered out by excluding the 
time periods where the count rate increased by $3\sigma$ from the quiescent 
state background rate. This resulted in net exposure times of
8.5ks and 13.1ks for the PN and MOS, respectively for the observation of May 2001. 
The observation of December 2001 was not affected severely by the particle 
induced flares: the net exposure times are 19.7ks for the PN and MOS cameras. 

The source spectra were extracted from the final filtered event lists using
 a circular region of radius $90\arcsec$ centered on the observed position of
 MCG-5-23-16 for both the PN and MOS cameras. Background spectra were extracted
 using appropriate annular regions around the observed position of MCG-5-23-16.
 Any source in the background region was masked out.
Appropriate response and effective area files for both the PN and MOS cameras
 were created using the SAS. 

\section{Spectral Analysis \label{analysis}} 
The PN and MOS spectra of MCG-5-23-16 were analyzed using the spectral fitting
 package {\tt XSPEC v11.2.0}. 
In order to check for  consistency in the spectral calibration of PN and MOS 
cameras, to check for spectral variability and to derive time averaged spectral
 characteristics of MCG-5-23-16, we have carried out spectral model fitting (i) to
 the individual PN and MOS data sets extracted from each observation, (ii) jointly to
 the PN and MOS data sets for each observation, and (iii) to the time averaged PN data
 obtained by combining the two PN data sets from the two observations. Thus, for
 spectral fitting, we formed nine data sets from the two observations carried out
 using the three imaging cameras: PN, MOS1, and MOS2.
The nine data sets  were first fitted by a simple absorbed power-law model over
 the entire $0.2-12\kev$ bandpass. None of the fits was acceptable e.g., the
 power-law fit to the combined PN spectrum resulted in a minimum $\chi^2$
 ($\chi^2_{min}$) of $2503.7$ for $504$ degrees of freedom (dof). 
To show the significant deviations from a best-fit power law and to determine the
 true continuum shape, we refitted the absorbed power-law model after excluding the
 significant features. For these fits we used the data in the $2.5-5\kev$ and 
$7.5-10\kev$ bands. In these bands, AGNs usually do not show any significant
 localized feature. Figure~\ref{f1} shows the combined PN data fitted with the 
absorbed power-law model and the ratio of data to the best-fit model
 extrapolated to lower and higher energies. As can be seen in Fig.~1, the poor
 fit is due to the significant features: a soft excess component below $\sim1\kev$,
 a strong iron K$\alpha$ emission line at $\sim 6.4\kev$, and likely excess
 emission above $10\kev$. This paper concentrates on the $2.5-12\kev$ band and the
 soft X-ray spectrum will be studied in a separate paper using RGS data also.
\subsection{The $2.5-10\kev$ continuum \label{continuum}}
To characterize the X-ray continuum, simple red-shifted power-law models modified
 by absorption due to an intervening medium at $z=0$ were fitted to the six spectra
 derived from the two observations with the three EPIC cameras: PN, MOS1, and MOS2. These fits 
were carried out over the $2.5-10\kev$ band excluding the FeK$\alpha$ region
 ($5-7.5\kev$). All the fits resulted in acceptable $\chi^2$; however, the best-fit
 values of equivalent hydrogen column ($\nh$) were much higher 
($\nh \sim 1.7\times10^{22}{\rm~cm^{-2}}$ for the PN data of December 2001) than
 the Galactic column of $8.82\times10^{20}{\rm~cm^{-2}}$ \citep{EWL89}.  Setting 
the $\nh$ parameter to the Galactic value resulted in unacceptable fits for all
 the data sets, thus implying heavy obscuration. The excess absorption column was
 then determined by introducing an additional absorption component at the source
 redshift and by refitting the power-law model to individual data sets. 
Table~\ref{tab1} lists the best-fit parameters: excess \nh, the photon index
 $\Gamma_X$, and the fit-statistic
$\chi^2_{min}$. All the errors quoted, here and below, are at the $90\%$ confidence
 level. Also listed in Table~\ref{tab1} are the observed flux in the $2.5-10\kev$ 
band and the absorption corrected (intrinsic) flux in the same band.  
As can be seen in Table~\ref{tab1}, the best-fit parameters \nh and $\Gamma_X$, 
derived from the data obtained simultaneously with different instruments, are 
consistent within the errors. However, the observed flux is found to differ by
 $\sim 20\%$ between EPIC PN and MOS cameras. This discrepancy is not unusual
 for X-ray instruments in their early phase of calibration.

In order to further constrain the spectral shape, the absorbed power-law model was
 fitted to the PN and MOS spectra jointly. The relative normalizations for the
 different instruments were kept free allowing for the small difference in the
 calibrated absolute flux, and any differences in the fraction of encircled counts 
contained in the PN and MOS extraction cells. The results of these fits are also
 listed in Table~\ref{tab1}. There are small variations in the best-fit spectral
 parameters between the two observations: $\Delta\nh = (4.4\pm2.7)\times10^{21}{\rm~cm^{-2}}$;
 $\Delta\Gamma_X=0.13\pm0.06$; $\Delta f_{int}=7\times10^{-12}\funit$. 
The source became harder and fainter in Dec. 2001.

To determine the average continuum shape, we fitted the absorbed and red-shifted
 power-law model to the time averaged PN data obtained by combining the two PN data
 sets from the two observations. The time averaged photon index is
 $\Gamma_X = 1.69\pm0.03$ and the observed flux in the $2.5-10\kev$ band is $7\times10^{-11}\funit$.

\subsection{The iron K$\alpha$ emission}
Figure~\ref{f2} shows the ratio of the PN spectrum extracted from the December 2001
 observation to the best-fit power law derived by using the data in the $2.5-5\kev$
 and $7.5-10\kev$ bands as described in the preceding section (see Table~\ref{tab1}).
 This plot clearly reveals a broad emission feature in the $5-7\kev$ band and
 peaking at $\sim6.4\kev$. This feature is common among Seyfert galaxies and is 
attributed to the fluorescence emission from the iron K-shell \citep[e.g.,][]{Nandraetal97}. The FeK$\alpha$ emission
 of MCG-5-23-16 appears to be nearly symmetric in its red and blue wings and is 
clearly different in shape from the red-shifted and asymmetric FeK$\alpha$ profiles
 generally observed from Seyfert~1 galaxies. Fig.~2 also suggests the likely
 presence of an iron K-edge at $\sim7.1\kev$.
  To measure the strength and shape of the FeK$\alpha$ emission, we parameterize
 the observed profile in terms of simple Gaussian models. For this analysis,  the
 co-added PN data were used, as the PN has the largest effective area among the three 
EPIC cameras. Hence the derived parameters characterize the time averaged FeK$\alpha$
 profile. Initially, a narrow unresolved Gaussian ($\sigma=0.01\kev$) was used
 for the FeK$\alpha$
 emission and the absorbed and red-shifted power-law model was used for the 
continuum as described in section~\ref{continuum} (Table~\ref{tab1}).  
This model resulted in a poor fit with $\chi^2_{min}$ of 262 for 235 dof. 
The ratio of the data to the best-fit narrow Gaussian model is plotted in
 Figure~\ref{f3} (second panel from the top) which shows a dip at $\sim6.4\kev$ 
and excess counts on the red and blue sides of the dip. This suggests a 
strong core at $\sim6.4\kev$ and broad wings on both sides of the core.
Varying the width of the Gaussian improved the fit significantly
 ($\Delta \chi^2 = 49$ for 1 additional parameter). This suggests that the
 line may be broad. Addition of a narrow Gaussian component ($\sigma=0.01\kev$) 
 at $\sim6.4\kev$ improves the fit significantly (($\Delta \chi^2 = 26.3$
 for two additional parameters). The best-fit parameters describing the
 observed FeK$\alpha$ profile are listed in Table~\ref{tab2}. The peak energies 
of both the narrow and broad components are consistent with neutral iron. The 
full width at half maximum (FWHM) of the broad component is $\sim 42000\kms$. 
Addition of an absorption edge at $7.1\kev$ results in marginal improvement in the fit 
($\Delta \chi^2 = 2.7$ for one additional parameter) at a significance level of 
$\sim 93\%$ based on an F-test \citep{Bevington69}.

The width of the broad iron K$\alpha$ line observed from MCG-5-23-16 is too 
large to be produced in regions other than the accretion disk around a 
super-massive black hole (SMBH). Therefore we checked whether the observed 
profile is consistent with a relativistic disk-line model \citep{Fabianetal89}.  
This model assumes a Schwarzschild geometry and the disk emissivity is a power-law
function of disk radius r i.e. $\propto r^{q}$. We fixed the inner disk radius
 at $6r_g$ and the outer disk radius at $500r_g$, where $r_g = GM/c^2$. First we 
fitted the disk-line model without a narrow Gaussian component. The continuum 
was the absorbed power-law model as before. The free parameters were the disk 
inclination angle ($i$) between our line of sight and the disk normal, the disk 
emissivity $q$, and the normalization of the disk line. This fit 
 resulted in an acceptable 
fit statistic ($\chi^2_{min} = 229.8$ for 236 dof). The ratio of the data 
and the best-fit model is plotted in Figure~{f4} (top panel).  
The dip seen at 
$\sim6.4\kev$ is due to the characteristic shape of the line profile from a
face-on disk -- namely a strong core and a red wing. 
 The disk inclination angle was found to be in the range $0^\circ-11^\circ$. 
This is physically inconsistent with a Seyfert~1.9 galaxy according to the 
standard unification scheme of Seyfert galaxies \citep{Antonucci93}. In view 
of the fact that the observed FeK$\alpha$ profile can be described by a 
combination of narrow and broad Gaussian and a narrow component is expected 
due to reflection from cold matter away from the disk, we added a narrow 
Gaussian component to our disk-line model. The energy of the narrow component
 was fixed at $6.4\kev$. This fit resulted in significant improvement over that
 without a narrow component ($\Delta\chi^2 = 32.4$ for one additional parameter).
 This is an improvement at a significance level of $>99.99\%$. The ratio of 
the data to the best-fit model is shown in Figure~\ref{f4} (second panel from 
the top). The best-fit inclination angle now is $i=46.3_{-3.8}^{+3.4}$. The 
equivalent width of the disk-line is $129_{-23}^{+31}\ev$ and the line energy 
is consistent with neutral iron. Addition of an absorption edge at $7.1\kev$
 does not lead to a significant improvement in the fit, and the upper limit to the 
optical depth is found to be $\tau < 0.03$. 

We have also tested the observed iron K$\alpha$ emission profile of 
MCG-5-23-16 against that expected from an accretion disk around a  
rotating (Kerr) black hole \citep{Laor91}. We fixed the inner radius 
at $1.23r_g$, the minimum radius allowed for a Kerr black hole,
and outer radius at $400r_g$ which is the 
maximum radius allowed by the model. As before, the disk emissivity is a 
power-law in radius ($\propto r^{-q}$). The variable parameters were 
$E_{FeK\alpha}$, $i$, $q$, and the line normalization. We used an absorbed 
and red-shifted power-law model for the continuum. The bottom two panels in
 Fig.~\ref{f4} show the ratio of the data and best-fit disk-line models for
 a Kerr black hole. Without a narrow Gaussian component, the fit is acceptable 
($\chi^2 = 251.7$ for 235) dof) but the best-fit inclination 
angle is $\sim 18\deg$. Inclusion of a narrow Gaussian component 
($\sigma=0.01\kev$) resulted in a significant improvement 
($\Delta \chi^2 = 60.5$ for one additional parameter) and in the best-fit 
inclination angle of $\sim47.4\deg$ (see Table~\ref{tab2}). Addition of an 
absorption edge at $7.1\kev$ does not improve the fit significantly 
($\Delta \chi^2 = 0.9$ for one additional dof).

In Figure~\ref{f5}, we have plotted the final best-fit models for the observed
iron K$\alpha$ line profile of MCG-5-23-16. The three models -- (i) a combination 
of a narrow Gaussian and broad Gaussian, (ii) a disk line for a Schwarzschild or 
(iii) a Kerr black hole all describe the observed data equally well.  

In order to look for any variations in the strength of the FeK$\alpha$ 
emission, we have also analyzed the observed FeK$\alpha$ profiles from 
the two \xmm{} observations. We used both PN and MOS data for this purpose. 
We fitted a combination of a broad and a narrow Gaussian profile to the PN and 
MOS data jointly for each observation. The $2.5-10\kev$ continuum was fixed as 
determined earlier (see Table~\ref{tab1}). The rest-frame line energies of both 
the narrow and broad components were fixed at $6.4\kev$. The results of these 
fits are listed in Table~\ref{tab3} for both the observations. It appears that 
the strength of the broad component of the FeK$\alpha$ line decreased by about a 
factor of two between the two observations taken six months apart,  while the 
narrow component did not vary significantly. 

\subsection{The Compton reflection model}
\begin{figure}
\includegraphics[width=10cm,angle=-90]{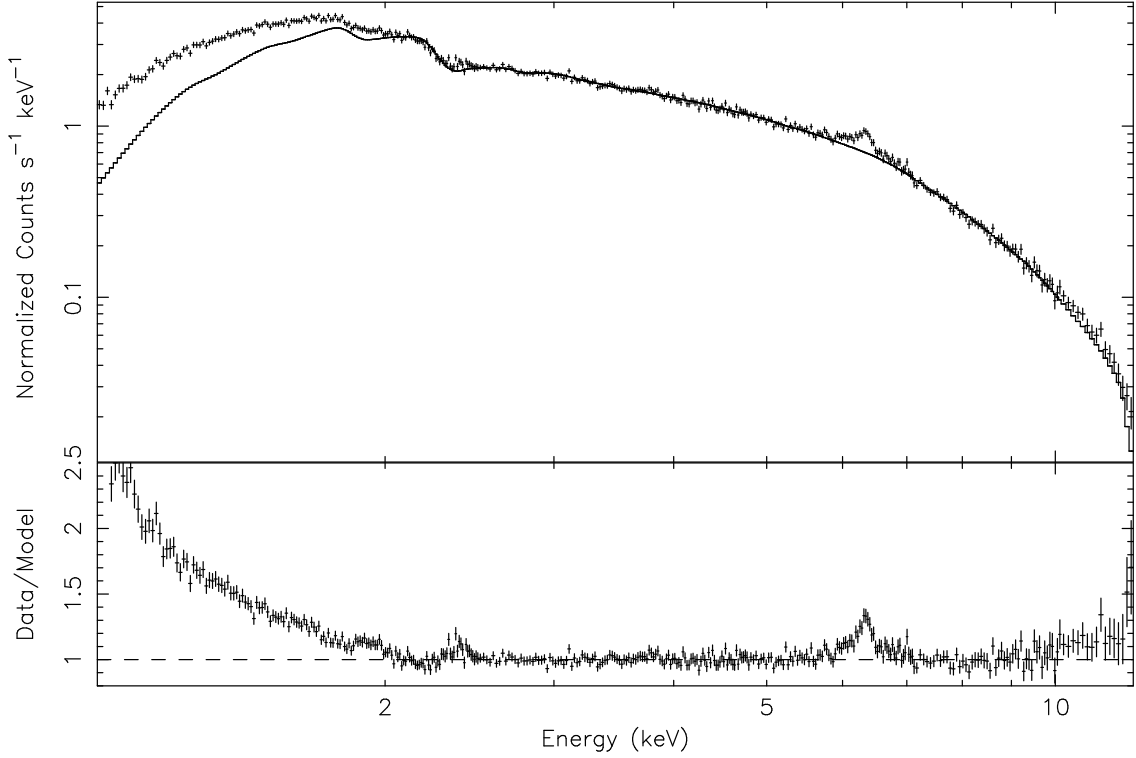} 
\caption{Ratio of combined PN data and the best-fit model derived by
fitting an absorbed and red-shifted power-law to the observed data
over the energy band of $2.5-10\kev$ but excluding the iron K$\alpha$
region $5.0-7.5\kev$ band. A strong soft X-ray excess below $2\kev$
and an emission feature at $\sim 6.4\kev$ due to iron K-shell emission
are seen clearly. Also seen is the excess emission over the power law
above $\sim10\kev$.}
 \label{f1}
\end{figure}

\begin{figure}
\includegraphics[width=10cm,angle=-90]{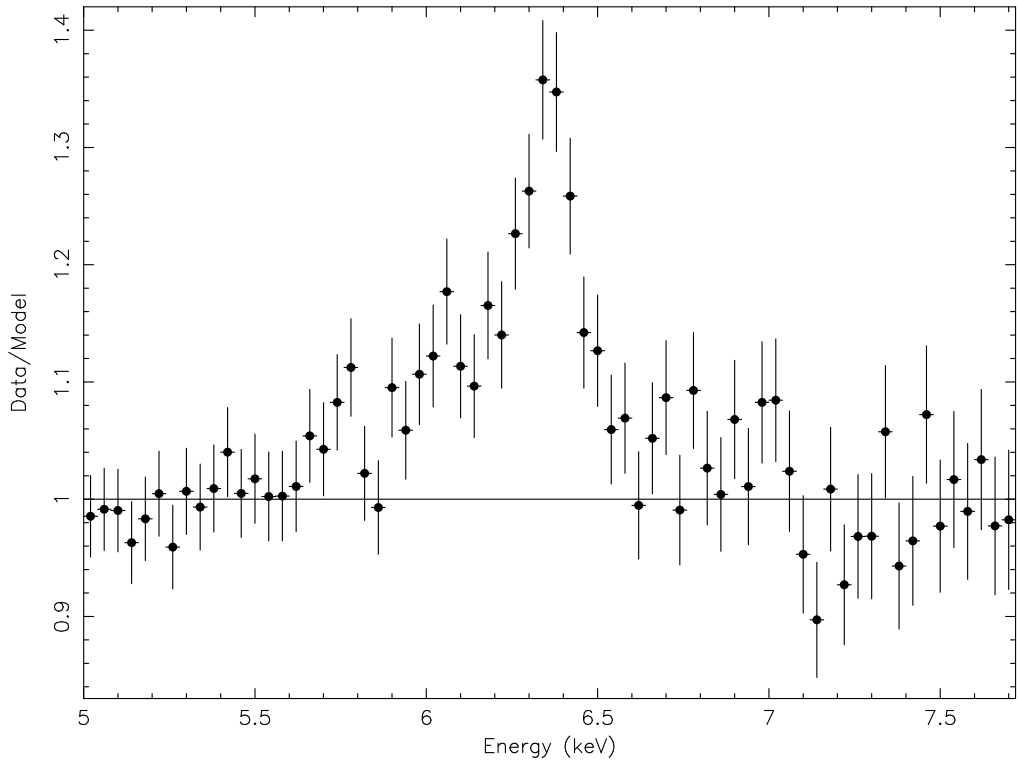}
\caption{The FeK$\alpha$ region of the X-ray spectrum of MCG-5-23-16 shown as the ratio of the observed data of December 2001 and the best-fit power-law model. 
The best-fit power-law continuum was derived by using the data in the $2.5-5\kev$
and $7.5-10\kev$ bands. A strong and broad FeK$\alpha$ peaking at
$\sim 6.4\kev$ is evident.
 \label{f2}}
\end{figure}

\begin{figure}
\begin{center}
\includegraphics[width=15cm,angle=-90]{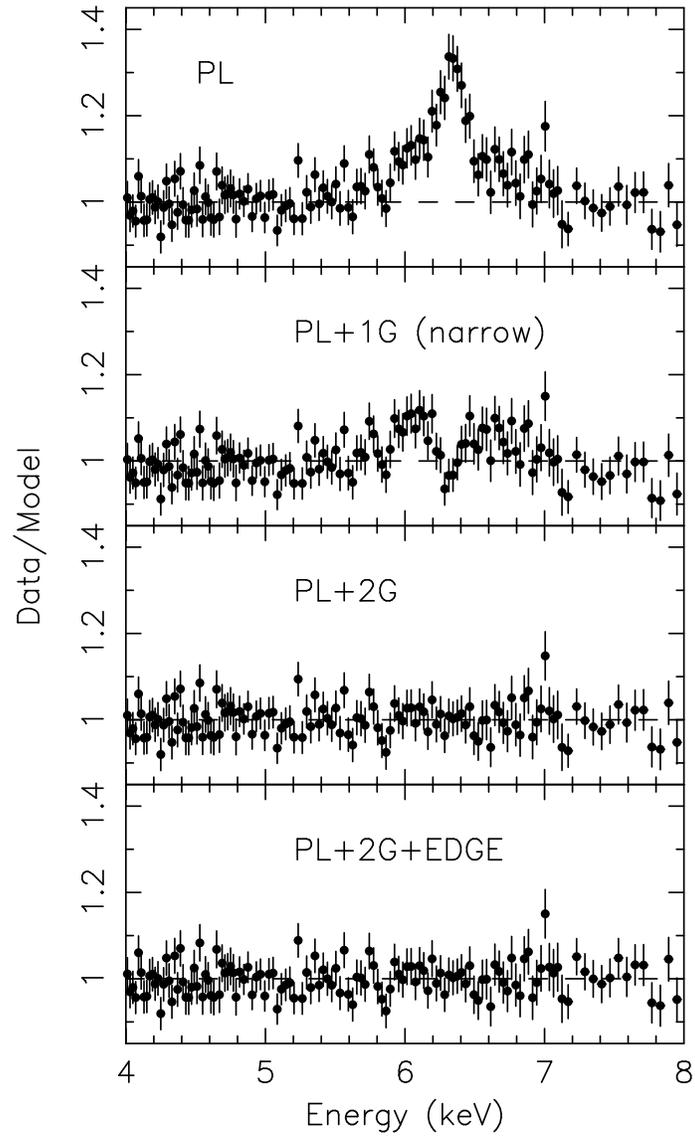}
\caption{Ratios of combined PN data to the best-fit Gaussian models of
the iron K$\alpha$ line profile. From top, the best-fit models are
(i) absorbed and red-shifted power law (PL), (ii) PL and a narrow Gaussian with
$\sigma=0.01\kev$, (iii) PL and a combination of a narrow and a broad
Gaussian, (iv) same as above but with an absorption edge at $7.1\kev$.}
\label{f3}
\end{center}
\end{figure}

\begin{figure}
\includegraphics[width=15cm,angle=-90]{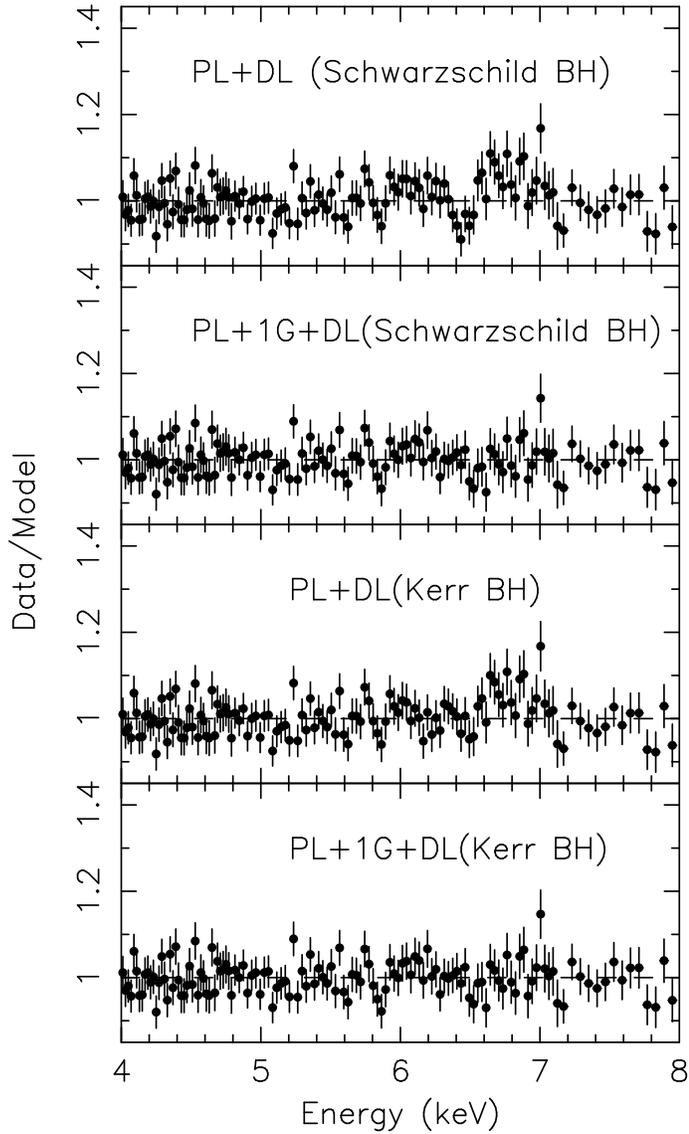}
\caption{Ratios of combined PN data to the best-fit relativistic
disk-line models for the iron K$\alpha$ line profile. From the top, the
best-fit models are (i) an absorbed and red-shifted power-law and a
disk-line model for a Schwarzschild black hole, (ii) same as above with an
addition of a narrow Gaussian line with $\sigma=0.01\kev$, (iii) PL and
disk-line model for a Kerr black hole, (iv) same as above with the addition
of a narrow Gaussian ($\sigma=0.01\kev$).}
\label{f4}
\end{figure}

\begin{figure}
\includegraphics[width=6cm,angle=-90]{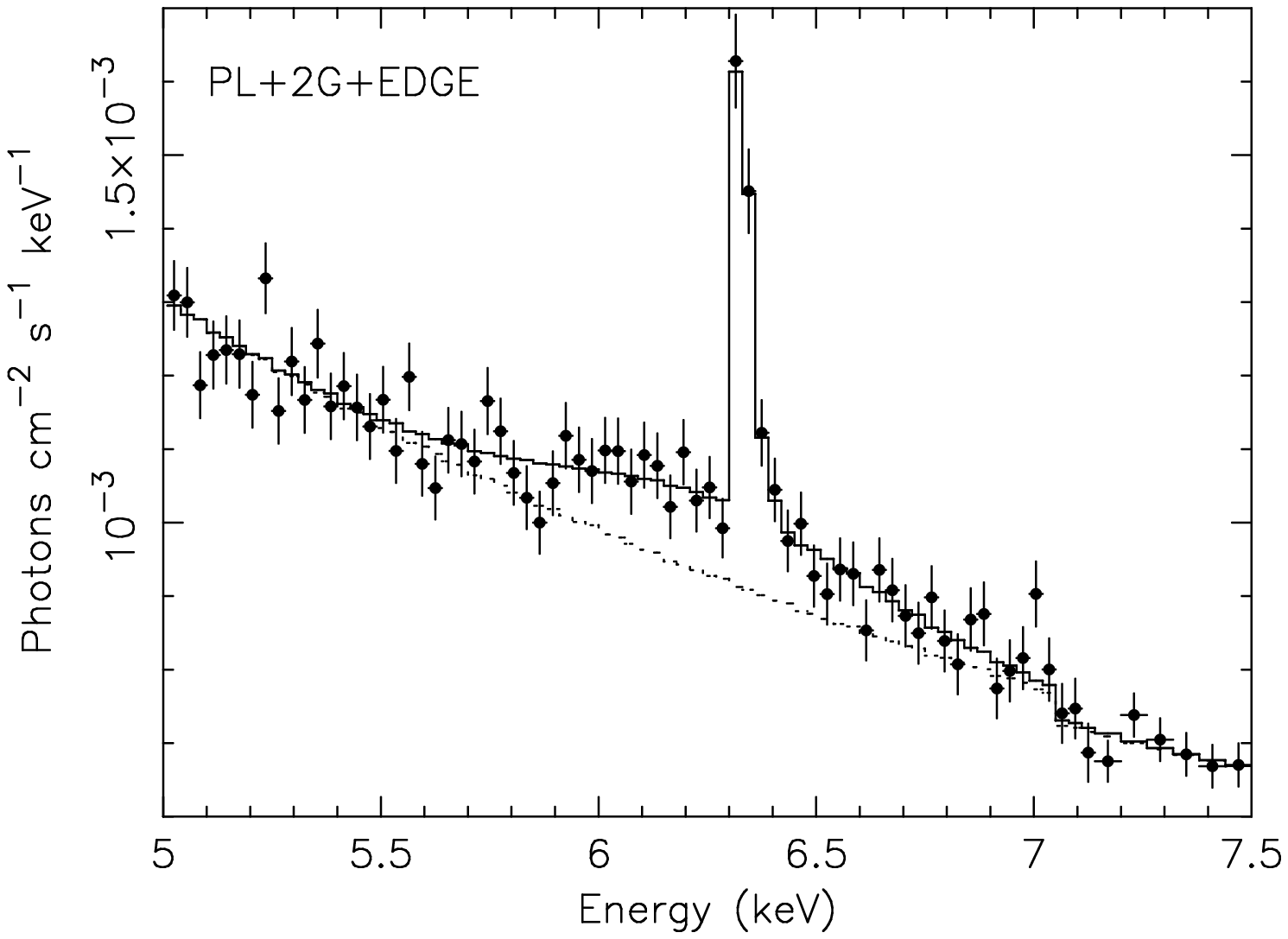}
\includegraphics[width=6cm,angle=-90]{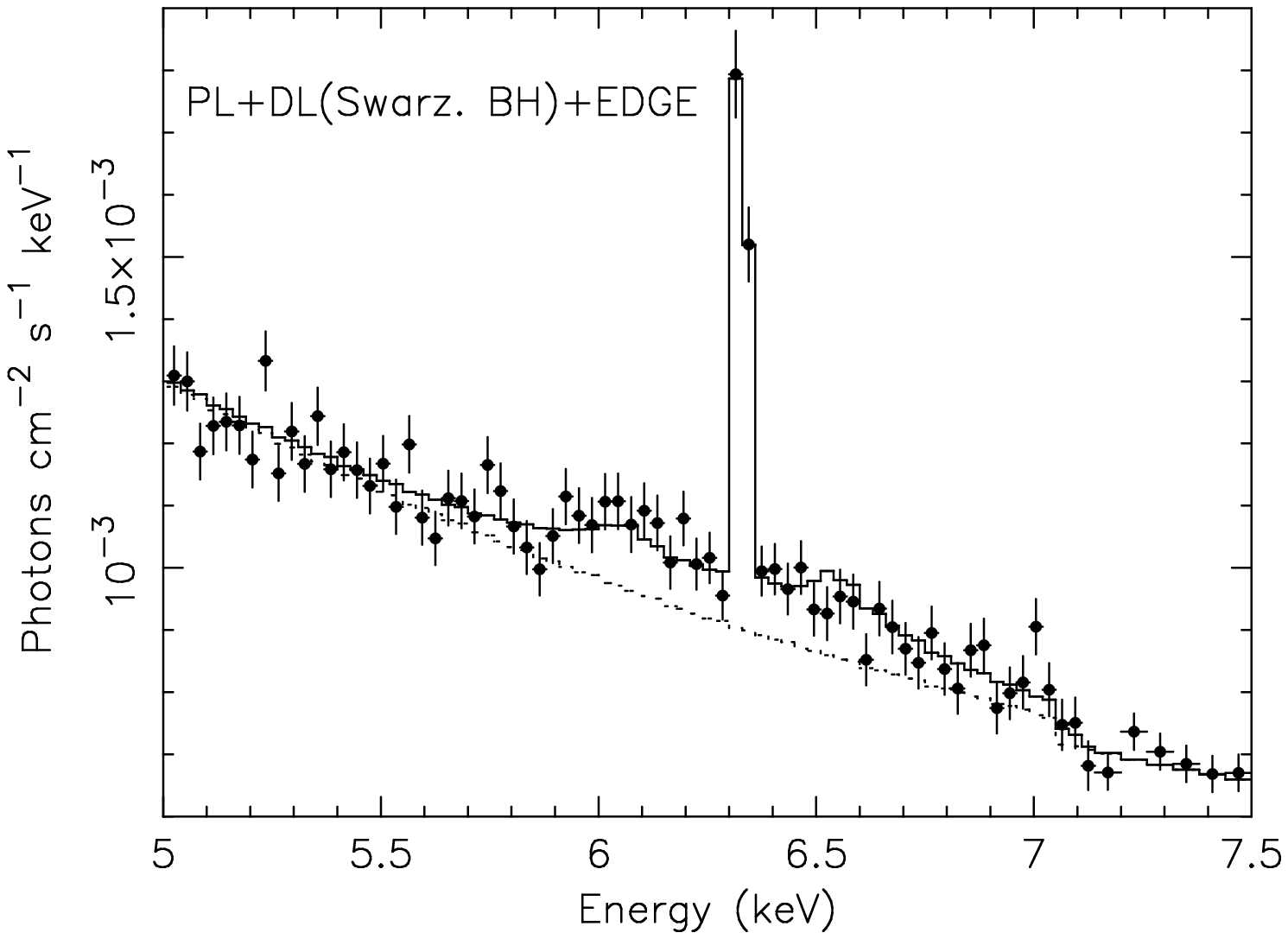}
\includegraphics[width=6cm,angle=-90]{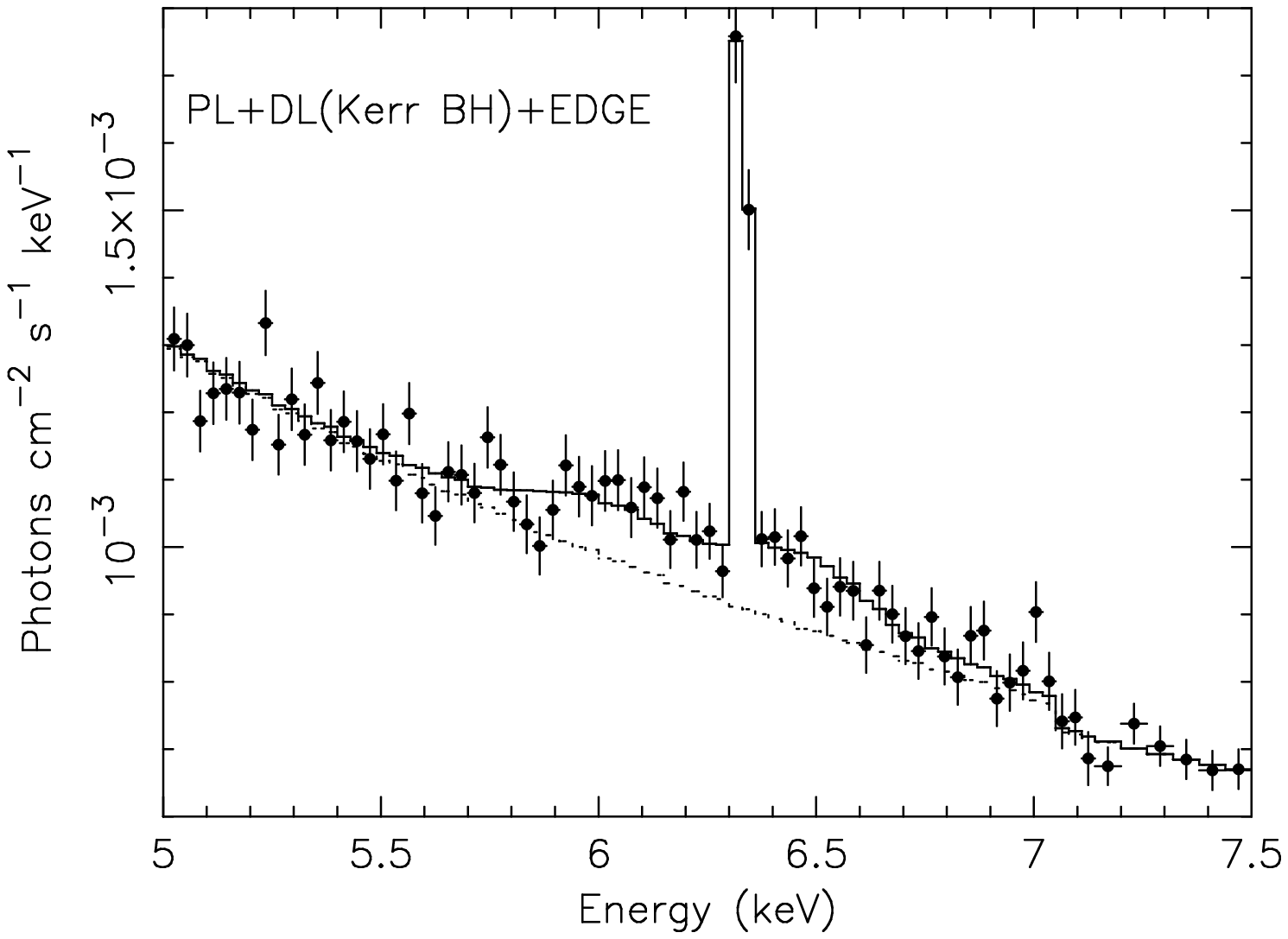}
\caption{The average EPIC PN FeK$\alpha$ line profile of MCG-5-23-16
and the best-fit model consisting of (i) a narrow Gaussian and a broad
Gaussian (top left panel), (ii) a combination of a narrow Gaussian and a
line arising from an accretion disk around a Schwarzschild black hole
(top right panel) or (iii) Kerr black hole (bottom panel).
 \label{f5}}
\end{figure}

\begin{figure}
\includegraphics[width=8cm,angle=-90]{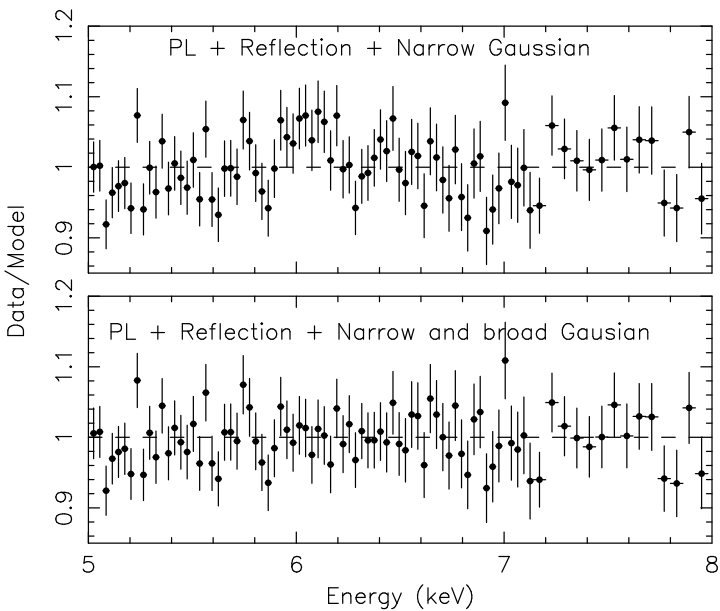}
\caption{Ratio of the combined PN data to the best-fit reflection
model with ($i$) a narrow Gaussian ($\sigma = 100\ev$) for the iron
K$\alpha$ emission (upper panel) and ($ii$) a narrow and a broad
Gaussian for the iron K$\alpha$ emission.
\label{f6}} 
\end{figure}

Under certain conditions, the Compton reflection emission can be important even 
below $\sim 10\kev$ (see section~\ref{orig_feka}). If the onset of the reflection 
component is at $5\kev$, the presence of a strong Fe~K-shell edge at $\sim 7.1\kev$ 
may artificially mimic a broad emission line feature at $\sim 6.4\kev$. This 
artificial ``broad line'' is likely to arise if the observed continuum is flatter 
and is dominated by the reflection component.
The $2.5-10\kev$ photon index of MCG-5-23-16 derived from the combined 
PN data ($\Gamma_X \sim 1.7$) is flatter than the average $2-10\kev$ 
photon index of Seyfert galaxies ($<\Gamma_X>\sim 1.9$; \citealt{Mushotzky97}). 
This flatter spectrum and strong iron K$\alpha$ line of MCG-5-23-16 could then be
partly  due 
to the  contribution of a Compton reflection process in a geometry where
 the primary X-ray source is partially covered by cold matter e.g. a torus. The 
reflection emission has been detected from MCG-5-23-16 with \rxte{} observations 
in 1995 \citep{WKP98}. To test whether the broad feature near $6.4\kev$ could 
be due to reflection and iron K-edge absorption of the primary emission, we have 
fitted a Compton reflection model to the combined PN data in the $2.5-12\kev$ band.
We used the  reflection model ``{\tt pexrav}'' \citep{MZ95} available with 
{\tt XSPEC}. This model calculates the reflected spectrum from a neutral disk 
exposed to an exponentially cut-off power-law spectrum \citep{MZ95}.  
Since \xmm{} does not cover the expected peak ($\sim30-40\kev$) of the reflection 
component, it is not possible to constrain all the parameters of the reflection
 model using \xmm{} data alone. Instead we fix the cut-off energy of the primary 
power-law at $200\kev$, disk inclination at $50\deg$, and the abundance of heavy
 elements at the solar value. The free parameters are the photon index of the 
primary power-law, the power-law normalization, and the relative amount of 
reflection compared to the directly viewed primary spectrum ($R$). We also 
included a narrow Gaussian line model to describe the FeK$\alpha$ 
emission. The results of this fit are listed in Table~\ref{tab4}. The best-fit 
photon index of the incident power-law continuum is $\sim1.8$ and is consistent 
with the mean photon index for Seyfert~1 galaxies. The ratio of the observed 
data to the best-fit model is shown in Figure~\ref{f6}. Although the fit 
is acceptable ($\chi^2_{min}=224.4$ for $269$ dof), there is weak excess emission 
near $6\kev$ suggestive of the possible presence of a weak broad emission line. To 
investigate further, we included an additional Gaussian component in our reflection
 model and carried out the fitting. This resulted in an improvement in the
 fit ($\Delta \chi^2=20.1$ at the cost of 2 dof). This is an improvement at a 
significance level of $99\%$ according to the maximum 
likelihood test. The best-fit parameters are listed in Table~\ref{tab4}.
The relative normalization of the reflection emission derived from the PN data is much higher than the value of $R=0.54_{-0.17}^{+0.26}$ inferred from the \sax{} observations of MCG-5-23-16 \citep{Risaliti02}. This is due to the excess emission above $\sim10\kev$ which amounts $\sim20\%$ above $10\kev$ (see Fig.~\ref{f1}). Such a large discrepancy in the reflection emission inferred from the PN and Bepposax data could be due to calibration uncertainties. This view is supported by MOS data, albeit lower signal-to-noise, which do not show a clear excess emission above $10\kev$. 

\begin{figure}
\includegraphics[width=8cm,angle=-90]{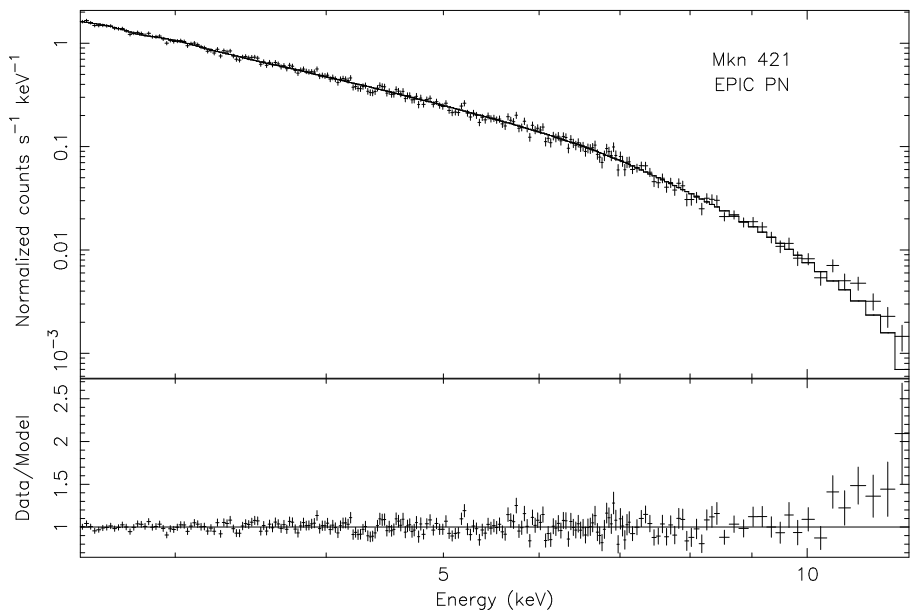}
\caption{The EPIC PN spectrum of Mkn~421 and the best-fit power-law model over the $2.5-12\kev$ band. A clear excess over the best-fit power-law is seen at energies above $\sim10\kev$. This excess is similar within error to that seen in the PN spectrum of MCG-5-23-16. \label{f7}}
\end{figure}

To test the reliability
of the excess emission above $10\kev$ in the \xmm{} PN data, we have 
analyzed the EPIC PN spectrum of a well
 known  BL Lac, Mkn~421. The BL Lac type AGNs do not show the Compton reflection
component, just a smooth continuum. Therefore, they are good targets to check
 the reliability of  spectral features in other AGN X-ray spectra.
We retrieved the data corresponding to the \xmm{} observation of Mkn~421 on 4
 May 2002 for an exposure time of $\sim40\ks$ from the public archive maintained
at HEASARC. The data were processed in a similar way as described above for
 MCG-5-23-16. The source spectrum was extracted from an annular region with inner
and outer radii of $10\arcsec$ and $80\arcsec$, respectively and centered at the
 observed source position. Due to the high flux of Mkn~421, the events from the
core of the point spread function were excluded to avoid possible pile-up which
results in artificial hardening of the spectrum. The background spectrum was
extracted from the source free regions near the position of Mkn~421. New
response files were created for Mkn~421. The PN spectrum was then grouped
appropriately and fitted by an absorbed power law model in the $2.5-10\kev$ band. Figure~\ref{f7} shows the PN spectrum of Mkn~421 fitted with a power-law and the ratio of data and the best-fit power-law model.
An  excess over the best-fit power-law is seen above $10\kev$. This excess is
$\sim20-30\%$ above $\sim10\kev$ which is similar to that seen in the PN spectrum of
 MCG-5-23-16 (Fig.~\ref{f1}). Since Mkn~421, being a blazar, is not expected
to show any excess emission, we conclude that the excess emission over the
power-law seen above $\sim10\kev$ in the spectra of Mkn~421 and MCG-5-23-16
is likely an artifact and reflects poor spectral calibration of the PN above
$10\kev$. Lack of any significant feature in the $2.5-10\kev$ band of the Mkn~421
  PN spectrum  suggests that the spectral calibration of PN in that
energy band is indeed good and the presence  of a broad iron K$\alpha$ emission
is not an artifact.

We refitted the Compton reflection model to the combined PN data after 
excluding the data above $10\kev$. We utilized the best-fit parameters for the 
reflection component derived from the \sax{} observations by \citet{Risaliti02} and  
fixed the cut-off energy $E_C$ at $157\kev$ and relative normalization $R$ at $0.54$. 
The best-fit parameters are listed in Table~\ref{tab4}. 
It is clear that a narrow Gaussian
and reflection emission cannot account for the strength of the broad feature 
at $\sim6.4\kev$ which implies the presence of a broad FeK$\alpha$ line.

\section{Discussion \label{discuss}}
We have presented the results from the two observations of MCG-5-23-16 using 
\xmm{}. The $2-10\kev$ flux  is found to be $6.5\times10^{-11}\funit$
 during December 2001 and $7\times 10^{-11}\funit$ during the June 2001 observation,
 comparable with the $7.3\times10^{-11}\funit$ measured during the \asca{}
 observations of 1994 \citep{Weaveretal97}. 
The $2.5-10\kev$ photon index appears to have flattened slightly 
from $1.77_{-0.05}^{+0.04}$ (May 2001) to $1.64_{-0.03}^{+0.03}$ (December 2001).
 Both these photon indices are steeper than the index of $1.4-1.5$ observed with
 \ginga when the source was in a lower flux state \citep{NP94,SD96}. 
 The X-ray spectrum of December 2001 is slight flatter than the \asca{} spectrum 
of 1994 when the source flux was slightly higher. These observations suggest 
that the X-ray spectrum flattens at lower flux levels -- a general trend well 
established for Seyfert galaxies \citep{SRV91,DMZ00,ZG01,VE01,Nandra01,Dewanganetal02}.

\subsection{FeK$\alpha$ emission and its variability}
\xmm{} observations of MCG-5-23-16 have reaffirmed the presence of a
complex iron K$\alpha$ emission earlier observed with \asca{}
\citep{Weaveretal97,WR98,Turneretal98} and \rxte{} \citep{WKP98}. Our
spectral fits imply that the line is broad, superimposed with a narrow
unresolved component, and possibly affected by an absorption edge of
iron at $7.1\kev$. The width of the broad line derived by fitting a
Gaussian profile suggests that this component could only be formed in
an accretion disk around a black hole. Indeed the broad component is
consistent with that expected from a relativistic accretion disk
around a Schwarzschild or a Kerr black hole. Our results also confirm
the \asca{} result that it is difficult to explain the entire line
profile as originating entirely in an accretion disk.  The Gaussian
and relativistic line model fits to the iron K$\alpha$ line imply two
distinct components. The narrow component is unresolved (FWHM $\la
1000\kms$) and has an equivalent width of $\sim 40\ev$. Similar narrow
unresolved components have been observed with \chandra{} and \xmm{}
from a number of AGNs
\citep[e.g.,][]{Kaspietal01,Poundsetal01,Turneretal02}, and are found
to have their peak energy at $6.4\kev$ with equivalent widths of $\sim
50\ev$ . Such narrow components are naturally expected from the torus
and/or the broad line region in the framework of the unification model of
Seyfert galaxies.  The maximum equivalent width of an FeK$\alpha$ line
produced by fluorescence in a cold reflector far from the disk with
$\nh \simeq 10^{23}{\rm~cm^{-2}}$ is $\simeq65\ev$ in the case of
NGC~4258 \citep{RNM00}. Thus the observed equivalent width of the
narrow iron F$\alpha$ line of MCG-5-23-16 is consistent with that
expected from the torus.

The broad component has an FWHM of $\sim 4\times10^4\kms$ and an
 equivalent width of $\sim110\ev$. This component is described by a
 broad Gaussian or relativistic disk-line arising from an accretion
 disk around a Schwarzschild or Kerr black hole suggesting that the
 line profile is not strongly modified by gravitational effects but
 broadened by the Doppler effect in a rotating disk. The disk-line fits
 imply that the disk is neutral and the disk-normal is inclined at an
 angle of $\sim50\deg$.  The likely presence of an absorption edge at
 $\sim 7.1\kev$ (see Fig.~\ref{f2}) also suggests that the disk is
 neutral.  The best-fit values of the emissivity indices inferred from
 the disk-line fits (in Schwarzschild and Kerr geometry) are similar
 but lower than the average value of $2.5$ obtained for Seyfert~1
 galaxies \citep{Nandraetal97}.  It should be noted that Nandra et
 al. had also fixed the inner radii to $6r_g$ (Schwarzschild black
 hole) and $1.23r_g$ (Kerr black hole). \citet{Weaveretal97} inferred
 $q$ to be in the range of $5-10$ but allowed $r_i$ to vary with the
 best-fit values in the range $9-22$. These results imply that if the
 emissivity law for MCG-5-23-16 is the same as that of an average
 Seyfert~1 galaxy, then the inner radius of the accretion disk is
 larger for MCG-5-23-16.

Our spectral fits suggest that the $2.5-10\kev$ continuum has flattened from
 $\Gamma_X \sim 1.77$ (May 2001) to $\Gamma_X \sim 1.67$ (December 2001) and the
 broad component of the iron K$\alpha$ line has weakened  from an equivalent
 width of $\sim180\ev$ (May 2001) to $\sim90\ev$ (December 2001). In the \asca{}
 observations of May 1991, the photon index and the equivalent width of the broad 
component of the iron K$\alpha$ line were found to be $1.79\pm0.08$ and 
$254_{-77}^{+55}$, respectively. These results suggest weakening of the broad 
iron K$\alpha$ line with flattening of the X-ray continuum.
Higher equivalent
 widths of the iron K$\alpha$ emission could be produced by an increased abundance
 of iron or by increasing the reflection component. It is unlikely that the
 change in the equivalent width of iron K$\alpha$ is due to a change in the 
iron abundance due to the following reasons: ($i$) it is very unlikely that
 iron abundance changes by a factor of about two in just six months, and ($ii$) 
it is difficult to explain the variation in the shape of the non-thermal X-ray
 continuum by changing the iron abundance.
Increasing the reflection emission from a neutral disk may flatten the continuum.
 This will require a change in the geometry of the accretion disk-corona system 
(e.g. increasing the covering factor at the corona will result in more reflection
 emission). However, this will increase the equivalent width of the iron K$\alpha$
 line with flattening of the continuum. Large equivalent widths of FeK$\alpha$ 
can be produced if the accretion disk is ionized \citep{Weaveretal97}. 
\citet{Dewangan02} reported a correlation between the shape of the X-ray 
spectrum and the energy of the FeK$\alpha$ line among radio-quiet type~1 AGNs 
and suggested that the accretion disks of AGNs with steeper continua are more 
ionized. The trend in $\Gamma_X$ and the equivalent width of iron K$\alpha$
 emission from MCG-5-23-16 could be understood if the shape of the X-ray continuum
 is coupled with the ionization stage of the disk material in MCG-5-23-16 in a similar
 way to that observed among the radio-quiet AGNs.  

\subsection{Is the broad feature at $\sim6.4\kev$ really an iron K$\alpha$ line from
 an accretion disk? \label{orig_feka}}
As shown in section~\ref{analysis}, the broad feature at $6.4\kev$ can be modeled
 as a broad Gaussian (FWHM$\sim40000\kms$) or a line profile arising from an 
accretion disk around a Schwarzschild or a Kerr black hole. Alternatively,
most of the flux in the broad component can also be accounted by reflection emission
.  However, the derived
 reflection fraction ($R\sim3$) is much larger than that inferred from the \sax{}
observations of MCG-5-23-16 ($R\sim 0.5$; \citet{Risaliti02}) and the average 
value for Seyfert~1 galaxies ($R \sim 1$; \citealt{Perolaetal02}). 
This is not entirely impossible
 if the reflector is a putative torus far away from the central continuum source.
 The reflection fraction can be large if the central source was brighter before
 a time period
which is equal to the time delay between the reflected and primary X-ray emission.
However, such a large reflection fraction for MCG-5-23-16 is unlikely due to the 
following reasons -- (i) the calibration of the EPIC PN above $10\kev$ is reliable at a level
of $\sim20\%$ (Snowden, private communication). MOS data do not show clear excess 
emission above $10\kev$ as is the case for PN. (ii) The broad feature at 
$\sim6.4\kev$ weakens with flattening of the continuum. This is not expected if the broad feature were the reflection emission. 
Simultaneous 
observations with \xmm{} and {\it Integral} would be crucial to characterize both the FeK$\alpha$ emission and reflection component.

\subsection{The Iron K-shell Edge}
\xmm{} observations of December 2001 have revealed the presence of an Iron K-shell
 absorption edge near $\sim 7.1\kev$. However, the edge is not well constrained
 and is detected only at a significance level of $93\%$. The edge energy is not
 well determined but it appears to be $\sim 7.1\kev$ (see Fig.~\ref{f2}). The
 optical depth is found to be $0.03_{-0.02}^{+0.04}$. Assuming that the edge 
arises due to the K-shell absorption of neutral iron and the corresponding 
photoionization cross section ($2.6\times10^{-20}{\rm~cm^{-2}}$;
 \citealt{Donnellyetal00}), we determine the column density of neutral iron 
to be $N_{\rm Fe} \approx 1.1\times10^{18}{\rm~cm^{-2}}$. This corresponds to 
an effective Hydrogen column density of $\nh \approx 4\times 10^{22}{\rm~cm^{-2}}$, 
which is a factor of two higher than the hydrogen column density derived by fitting
 an absorbed power-law model to the observed data. Given the uncertainty in the 
derived value of the optical depth, it cannot be ruled out that the Fe K-edge 
arises entirely in the absorbing material along our line of sight in the form
 of a putative torus. 
It is interesting to investigate whether both the absorption K-edge and the
 narrow component of the K$\alpha$ line due to iron could be produced in the
 the putative torus around an accretion disk. In the case of isotropic emission,
  the equivalent width of the emission line should be 
 $\sim 64f_c \nh A_{Fe} \times 10^{-23}$.
 For a typical covering factor $f_c\sim0.7$ for the torus and solar abundances 
($A_{fe} = 1$), the absorption column required to produce an equivalent width of
  $40\ev$ for the narrow iron K$\alpha$ line is $ \sim 10^{23}{\rm~cm^{-2}}$.
 This column is larger by a factor of five than the absorption column derived
 from the power-law fit. It is also larger by about a factor of three than the
 absorption column derived from the iron K-edge, since the absorption features
 are produced only by  a column along our line of sight while the emission
 features are spatially integrated in the case of a source which is unresolved.
 The above results imply that we are viewing the outer edge of a putative torus 
where the column density is lower than the inner regions of the torus. This is
 consistent with that expected for a Seyfert~1.9 galaxy from the unification 
scheme \citep{Antonucci93}.
\section{Conclusions \label{conclusion}}
We studied the $2.5-10\kev$ spectrum of MCG-5-23-16 which was observed twice
 with \xmm{} in May and December 2001. Our main results are as follows. \\
\begin{enumerate}
\item The iron K$\alpha$ line profile of MCG-5-23-16 is broad and nearly symmetric.
 The observed profile could be modeled by a combination of a narrow (unresolved) and
 a broad Gaussian component.
 The broad component is also consistent with that expected from an
 accretion disk around a Schwarzschild or a Kerr black hole.
\item The disk inclination angle of $\sim 46^\circ$, inferred from the disk-line
 fits, is consistent with that expected for a Seyfert~1.9 galaxy according to the
 Seyfert unification picture.
\item Alternatively, most of the flux of the broad feature at $6.4\kev$ can also be modeled as 
 the reflection emission, however, the reflection fraction ($\sim3$) is much larger than that derived from the \sax{} observations ($R\sim 0.5$). This is likely to be due to uncertainty in the calibration of EPIC PN above $10\kev$.
\item We have derived three values of absorption column 
($i$) $\nh \sim 10^{22}{\rm~cm^{-2}}$ from the best-fit absorbed and red shifted 
power-law model, ($ii$) $\nh\sim4\times10^{22}{\rm~cm^{-2}}$ from the absorption
 edge, and ($iii$) $\nh\sim 10^{23}{\rm~cm^{-2}}$ required to produce the narrow
 iron K$\alpha$ line emission. The above three values are consistent with a 
geometrical picture in which our line of sight passes through the outer edges
 of a putative cold torus obscuring the nuclear X-ray source as expected for
 a Seyfert~1.9 galaxy from the unification scheme. 
\item The broad iron K$\alpha$ emission appears to weaken with the flattening of
 the X-ray continuum. This can be understood if the degree of ionization of iron
 in the accretion disk decreases with flattening in the X-ray spectrum. However,
in any case the disk does not appear to be highly ionized 
 (i.e. Hydrogen or Helium-like iron). 
\end{enumerate}

\acknowledgements
This work is based on observations obtained with \xmm{}, an ESA science mission 
with instruments and contributions directly funded by ESA member states and the
 USA (NASA). The authors thank the referee of this paper, Dr. Kim Weaver, for her comments and suggestions. REG acknowledges NASA award NAG5-9902 in support of his 
Mission Scientist position on \xmm{}.

\clearpage

\clearpage
\begin{table}
\caption{Best-fit absorbed power-law model parameters for MCG-5-23-16 \label{tab1}}
{\small
\begin{tabular}{lccccccc}
\tableline\tableline
Observation & Data\tablenotemark{a}  & Model\tablenotemark{b} & N$_{\rm H}$\tablenotemark{c}& $\Gamma_{\rm X}$ & $f_obs$ & $f_int$ & $\chi^2/dof$ \\
            &       &       & $10^{22}{\rm~cm^{-2}}$   &            &    &  &  \\
\tableline
2001 December      &  PN  & PL & $1.67_{-0.19}^{+0.19}$ & $1.66_{-0.03}^{+0.03}$ & 6.9 & 7.6 & $138.4/146$  \\
       & MOS1 & PL & $1.23_{-0.42}^{+0.42}$ & $1.52_{-0.09}^{+0.09}$ & 5.7 & 6.1 & 169.8/161 \\
       & MOS2 & PL & $1.56_{-0.43}^{+0.43}$ & $1.55_{-0.09}^{+0.10}$ & 5.5 & 6.0 & 166.2/161    \\
 &     PN+MOS1+MOS2 & PL & $1.65_{-0.15}^{+0.16}$ & $1.64_{-0.03}^{+0.03}$ & 6.0 & 6.5 & 478.0/472 \\

2001 May      & PN  & PL &  $1.93_{-0.28}^{+0.28}$ & $1.75_{-0.04}^{+0.04}$ & 7.1 & 7.9  & 160.4/146  \\
        & MOS1 & PL & $2.03_{-0.48}^{+0.48}$ & $1.69_{-0.10}^{+0.10}$ & 6.4 & 7.2 & 182.9/161  \\
       & MOS2 & PL & $1.89_{-0.48}^{+0.48}$ & $1.70_{-0.11}^{+0.11}$ & 6.2 & 6.9 & 195.7/161   \\
      &  PN+MOS1+MOS2  & PL & $2.09_{-0.21}^{+0.22}$ & $1.77_{-0.05}^{+0.04}$ & 6.4 & 7.2 & 553/472 \\
2001 Dec. + May & PN & PL & $1.76_{-0.17}^{+0.17}$ & $1.69_{-0.03}^{+0.03}$ & 7.0 & 7.7 & 123.6/158 \\
\tableline 
\end{tabular}}
\tablenotetext{a}{The data in the $2.5-10\kev$ were used.}
\tablenotetext{b}{Simple red-shifted power-law modified by absorption}
\tablenotetext{c}{Excess absorption intrinsic to the source apart from that due to the Galactic column of $8.82\times10^{-20}{\rm~cm^{-2}}$.}
\end{table}

\begin{table}
\caption{Best-fit parameters of the two component Gaussian model for the iron K$\alpha$ emission of MCG-5-23-16 \label{tab2}}
{\footnotesize
\begin{tabular}{lcccccc}
\tableline\tableline
Parameter & \multicolumn{6}{c}{PL} \\ \tableline
         &\multicolumn{2}{c}{PN+MOS} & \multicolumn{2}{c}{PN+MOS} & \multicolumn{2}{c}{Combined PN} \\
         & \multicolumn{2}{c}{Dec. 2001} &  \multicolumn{2}{c}{May 2001}  &  \multicolumn{2}{c}{May and Dec. 2001} \\
         & NO EDGE & EDGE & NO EDGE  & EDGE &  NO EDGE & EDGE      \\
\tableline
$\Gamma_X$ & 1.64 & 1.64 & 1.77 & 1.77 & $1.69_{-0.04}^{+0.03}$ & $1.65_{-0.06}^{+0.05}$  \\
$\nh$      & 1.65 & 1.65 & 2.09 & 2.09 & $1.75_{-0.16}^{+0.16}$ & $1.60_{-0.22}^{+0.22}$ \\
$E_N$      & 6.4  & 6.4  & 6.4 & 6.4 &   6.4 & $6.39_{-0.03}^{+0.02}$ \\
$\sigma_N$ & 0.01  & 0.01  & 0.01 &  0.01 &  0.01 & 0.01  \\
$EW_N$     & $48.0_{-10.2}^{+7.8}$ & $52.3_{-14.5}^{+6.6}$ &$49.1_{-16.0}^{+13.2}$ & $46.1_{-11.0}^{+11.2}$ & $35.3_{-9.6}^{+9.6}$ & $35.2_{-10.0}^{+9.6}$ \\
$E_B$      & $6.4$ & $6.4$ & 6.4 & 6.4 & $6.32_{-0.08}^{+0.07}$ & $6.32_{-0.09}^{+0.09}$  \\
$\sigma_B$ & $0.370_{-0.076}^{+0.109}$ & $0.376_{-0.074}^{+0.143}$ & $0.550_{-0.110}^{+0.132}$ & $0.602_{-0.138}^{+0.182}$ & $0.381_{0.087}^{+0.104}$ & $0.355_{-0.060}^{+0.128}$  \\
$EW_B$     & $91.5_{-12.9}^{+21.5}$ & $92.6_{-14.2}^{+18.4}$ & $177_{-30}^{+38}$ & $189_{-37}^{+44}$ & $114_{-21.4}^{+21.0}$ & $96.4_{-30.7}^{+31.6}$ \\
$E_{edge}$ & -- &  7.1 & -- & 7.1 & -- & 7.1  \\
$\tau$     & -- & $0.02_{-0.02p}^{+0.02}$ & -- & $0.02_{-0.02p}^{+0.03}$ & -- & $0.03_{-0.02}^{+0.04}$ \\
$\chi^2_{min}$ & 727.2 & 722.9 & 783.8 & 782.5 &  186.7 & 184.0 \\
dof &  697 & 696 & 703 & 702 & 232 & 231 \\
\tableline
\end{tabular}}
\end{table}

\begin{table}
\caption{Results for relativistic disk-line fits to the combined PN data \label{tab3}}
\begin{tabular}{lcccc}
\tableline\tableline
Parameter & \multicolumn{2}{c}{Schwarzschild geometry} & \multicolumn{2}{c}{Kerr geometry} \\
          & NO EDGE & EDGE  & NO EDGE & EDGE \\
\tableline
$\Gamma_X$ & $1.68\pm0.02$ & $1.68\pm0.02$ & $1.68\pm0.03$ & $1.65_{-0.10}^{+0.05}$ \\
Normalization\tablenotemark{a} & $0.021\pm0.001$ & $0.021\pm0.001$ & $0.021\pm0.001$ & $0.021\pm0.001$ \\
$\nh$($\times10^{22}{\rm~cm^{-2}}$) & $1.72_{-0.16}^{+0.12}$ & $1.72_{-0.17}^{+0.14}$ & $1.72_{-0.21}^{+0.18}$ & $1.62_{-0.17}^{+0.17}$  \\
$E_N$($\kev$) & 6.4 & 6.4 & 6.4 & 6.4 \\
$\sigma_N$($\ev$) & 100 & 100 & 100 & 100 \\
$EW_N$($\ev$) & $43.5_{-9.6}^{+9.6}$ & $44.7_{-7.0}^{+7.4}$ & $38.0_{-9.6}^{+9.3}$ & $38.5_{-9.6}^{+9.3}$ \\
$E_D$($\kev$) & $6.31_{-0.25}^{+0.06}$ & $6.30_{-0.23}^{+0.07}$ & $6.29_{-0.12}^{+0.09}$ & $6.30_{-0.12}^{+0.09}$ \\
$q$ &  $1.81_{-0.37}^{+0.40}$ & $1.81_{-0.38}^{+0.41}$ & $1.75_{-0.55}^{+0.38}$ & $1.73_{-0.61}^{+0.36}$ \\
$r_i/r_g$ & 6 & 6 & 1.23 & 1.23 \\
$r_o/r_g$ & 400 & 400 & 400 & 400 \\
$i$ &  $46.3_{-3.8}^{+3.4}$ & $46.0_{-3}^{+2.5}$ & $47.4_{-6.7}^{+6.9}$ & $47.9_{-7.1}^{+9.9}$ \\
$EW_D$($\ev$)  &  $119_{-23}^{+31}$ & $128_{-22}^{+23}$ & $118_{-25}^{+28}$ & $116_{-25}^{+27}$ \\
$E_{edge}$($\kev$) & -- & 7.1 & -- & 7.1  \\
$\tau$ & -- & $0.010_{-0.0p}^{+0.021}$ & -- & $0.03_{-0.03p}^{+0.08}$ \\
$\chi^2/dof$ & 190.7/232 & 190.4/231 & 191.1/232 & 188.7/231 \\
\tableline
\end{tabular}
\tablenotetext{a}{Power-law normalization at $1\kev$ in units of ${\rm photons~\kev^{-1}~s^{-1}~cm^{-2}}$.}
\end{table}

\begin{table}
\caption{Best-fit parameters of a reflection model for the continuum and Gaussian model for the iron K$\alpha$ emission of MCG-5-23-16 \label{tab4}}    
\begin{tabular}{lcccc} 
\tableline\tableline 
 Parameter & \multicolumn{4}{c}{PL + Reflection} \\
           & \multicolumn{2}{c}{Reflection parameters varied} & \multicolumn{2}{c}{Reflection parameters fixed\tablenotemark{a}} \\
           & \multicolumn{1}{c}{Single Gaussian} & \multicolumn{1}{c}{Double Gaussian} & \multicolumn{1}{c}{Single Gaussian} & \multicolumn{1}{c}{Double
 Gaussian} \\ 
\tableline 
$\nh$ ($\times10^{22}{\rm~cm^{-2}}$) & $1.8_{-0.2}^{+0.1}$ & $1.75_{-0.08}^{+0.06}$ & $1.75_{-0.15}^{+0.08}$ & $1.67_{-0.05}^{+0.08}$ \\
$\Gamma_X$ & $1.85_{-0.05}^{+0.03}$ & $1.78_{-0.02}^{+0.02}$ & $1.68_{-0.03}^{+0.03}$ & $1.68_{-0.02}^{+0.05}$  \\ 
$E_{fold}$ & $200\kev$ & $200\kev$ & $157\kev$ & $157\kev$ \\
$R$ & $2.5_{-0.4}^{+0.3}$ & $1.7_{-0.2}^{+0.2}$ & $0.54$ & $0.54$ \\ 
$i$ & $60^\circ$ & $60^\circ$ & $60^\circ$ & $60^\circ$ \\   
$E_N$ & $6.37_{-0.02}^{+0.03}$ & $6.4$ & $6.4$ & $6.4$ \\
$\sigma_N(\ev)$ & $0.01$ & $0.01$ & $0.01$ & $0.01$ \\   
$EW_N$ ($\ev$) & $52.0_{-7.7}^{+6.9}$ & $43.2_{-11.4}^{+6.1}$ & $62.0_{-6.4}^{+9.3}$ & $34.5_{-11.4}^{+10.2}$ \\ 
$E_B$ & -- & $6.20_{-0.12}^{+0.10}$ & -- & $6.28_{-0.08}^{+0.08}$ \\    
$\sigma_B$ & -- & $0.18_{-0.12}^{+0.09}$ & -- & $0.31_{0.09}^{+0.12}$ \\
$EW_B$ & -- & $29.2_{-13.8}^{+33.5}$ & -- & $83.9_{-22.6}^{+12.3}$ \\
$\chi^2_{min}$ & 224.4 & 204.3 & 234.6 & 184.3\\
$dof$ & 269 & 267 & 236 & 233 \\
\tableline 
\end{tabular}
\tablenotetext{a}{Relative reflection $R$ and cut-off energy $E_C$ were fixed at the values derived from the \sax observations \citep{Risaliti02}.}
\end{table}

\end{document}